**Key Points:**
- Identifies substorm expansion, recovery, and possible growth phases from magnetic indices
- Phases identified from percentiles of the index rate change not fixed values
- Technique provides greater information about substorms than existing lists

**Supporting Information:**
- Figure S1
- Figure S2
- Figure S3
- Data Sets S1–S3

**Correspondence to:**
C. Forsyth,
colin.forsyth@ucl.ac.uk

# A new technique for determining Substorm Onsets and Phases from Indices of the Electrojet (SOPHIE)

**C. Forsyth[1], I. J. Rae[1], J. C. Coxon[2], M. P. Freeman[3], C. M. Jackman[2], J. Gjerloev[4,5], and A. N. Fazakerley[1]**

[1]UCL Mullard Space Science Laboratory, Dorking, UK, [2]Department of Physics and Astronomy, University of Southampton, Southampton, UK, [3]British Antarctic Survey, Cambridge, UK, [4]Johns Hopkins University Applied Physics Laboratory, Laurel, Maryland, USA, [5]Birkeland Centre of Excellence, University of Bergen, Bergen, Norway

**Abstract** We present a new quantitative technique that determines the times and durations of substorm expansion and recovery phases and possible growth phases based on percentiles of the rate of change of auroral electrojet indices. By being able to prescribe different percentile values, we can determine the onset and duration of substorm phases for smaller or larger variations of the auroral index or indeed any auroral zone ground-based magnetometer data. We apply this technique to the SuperMAG AL (SML) index and compare our expansion phase onset times with previous lists of substorm onsets. We find that more than 50% of events in previous lists occur within 20 min of our identified onsets. We also present a comparison of superposed epoch analyses of SML based on our onsets identified by our technique and existing onset lists and find that the general characteristics of the substorm bay are comparable. By prescribing user-defined thresholds, this automated, quantitative technique represents an improvement over any visual identification of substorm onsets or indeed any fixed threshold method.

## 1. Introduction

Substorms are the elemental dissipative events in the coupled solar wind-magnetosphere-ionosphere system that process ~$10^{15}$ J of captured solar wind energy during their lifetime [*Tanskanen et al.*, 2002]. In the magnetosphere, substorm expansion phases are accompanied by a dipolarization of the magnetotail magnetic field, an injection of energetic particles into the inner magnetosphere, a reduction of the magnetic flux within the magnetotail lobes, and a diversion of the cross-tail current into the ionosphere. In the ionosphere, substorm expansion phases are accompanied by a brightening and expansion of the nightside aurora [*Akasofu*, 1964], electromagnetic ULF waves (for a review, see Rae, I.J. and Watt, C.E.J., ULF waves above the nightside auroral oval during substorm onset, accepted in AGU Geophysical Monograph Series) and an enhancement in the auroral electrojet current due to the cross-tail current diversion, which results in a deflection of the magnetic field at ground level.

Substorms are typically broken down into three phases: growth, expansion, and recovery. During the growth phase, first identified by *McPherron* [1970] and which, on average, lasts 30–90 min [*Li et al.*, 2013], magnetic flux is added to the magnetotail lobes through reconnection at the dayside magnetopause. This process enhances magnetospheric convection [*Axford*, 1969] and the ionospheric electrojets, resulting in a small deflection of the *H* component of the ground magnetic field at auroral latitudes [*McPherron*, 1970]. As lobe magnetic flux increases, the auroral oval moves equatorward [*Coumans et al.*, 2007] and the temperature of the plasma sheet increases [*Forsyth et al.*, 2014]. At the onset of the expansion phase (often referred to as the substorm onset) there is an exponential increase in the auroral intensity [*Voronkov et al.*, 2003] and ULF wave activity [*Voronkov et al.*, 2003; *Rae et al.*, 2012] and the aurora expand poleward over ~15 min [*Partamies et al.*, 2013; *Chu et al.*, 2015]. In the magnetosphere, magnetotail currents are diverted into the ionosphere through field-aligned current systems [*McPherron et al.*, 1973], enhancing the westward electrojet and resulting in a formation of sharp negative bays in the *H* component of the ground magnetic field [*Akasofu and Chapman*, 1961; *Davis and Sugiura*, 1966]. The end of the expansion phase and start of the recovery phase are indicated by a reduction of the auroral intensity, ULF wave activity, and strength of the westward electrojet currents. Over ~1 h, the magnetospheric current systems are re-organized and distinct auroral features, such as "omega bands" and pulsating auroral patches, are observed [*Opgenoorth et al.*, 1994]. While isolated substorms usually follow this growth-expansion-recovery paradigm, events with multiple onsets or intensifications, seen as expansion phases occurring immediately following a recovery phase, are also reported [*Pytte et al.*, 1976].

In order to determine these and other repeatable physical processes during substorms, lists of substorm onsets have been generated using data from space- and ground-based auroral imagers and ground-based magnetometers [*Liou et al.*, 2001; *Frey et al.*, 2004; *Nishimura et al.*, 2010; *Newell and Gjerloev*, 2011]. These lists have then formed the basis of statistical studies, such as superposed epoch analyses [e.g., *Kistler et al.*, 2006; *Gjerloev et al.*, 2007; *Boakes et al.*, 2009; *Milan et al.*, 2009].

There is a clear link between auroral brightenings at substorm onset and enhancements in the westward electrojet that deflect the *H* component of the ground magnetic field [*Heppner*, 1954; *Akasofu*, 1964]. As such, auroral indices which combine data from magnetometers around the auroral zone are a useful alternative to direct auroral observations for determining substorm onset since these magnetometer data sets are almost continuously available over long periods of time. Specifically, the *AL* index [*Davis and Sugiura*, 1966] and SuperMAG AL (SML) index [*Newell and Gjerloev*, 2011; *Gjerloev*, 2012] act as virtual magnetometer stations that track the peak of the westward auroral electrojet irrespective of latitude or local time differing in the number and latitudinal extent of contributing stations. Yearly means of the auroral indices *AU*, *AL*, and *AE* follow the solar cycle, maximizing in the declining phase [*Ahn et al.*, 2000], and the *AL* and *AE* auroral indices have bimodal lognormal distributions [*Vassiliadis et al.*, 1996], with the two distributions being described as quiet or laminar and disturbed or turbulent. In contrast, the distributions of fluctuations in *AE* are smoothly varying [*Consolini and De Michelis*, 1998]. As one would expect, SML has similar distributions (shown in the supporting information).

Given the well-defined magnetic field profile of the expansion and recovery phases in auroral zone magnetic indices, individual substorm phases can also be readily identified. Here we propose a quantitative technique that not only determines substorm expansion phase onsets but also identifies the onset and duration of each individual substorm phase using auroral indices. We demonstrate the advantages of using this technique by applying it to the SML index. We compare the substorm expansion phase onset times derived using our technique with previously published onset lists using other methods.

## 2. Previous Methods for Determining Substorm Onset

### 2.1. Visual Identification of Auroral Onset

The most commonly referred to techniques for identifying substorm onset hark back to the definition by *Akasofu* [1964]—"a sudden increase in the brightness of… a quiet arc and subsequent rapid motion of the arc towards the geomagnetic pole." Using this, substorm onsets have been visually identified from space-based auroral imagers [*Liou et al.*, 2001; *Frey et al.*, 2004; *Frey and Mende*, 2006] and ground-based all-sky cameras [*Nishimura et al.*, 2010]. However, these methods are subjective, depending on the observer's judgment as to when an exponentially growing auroral arc started to brighten and whether the aurora expanded, and indeed how far. As a consequence, the results of these studies are nonrepeatable and cannot be applied to different data sets. Measures that are quantitative and objective measures are more useful [*Murphy et al.*, 2009b].

### 2.2. Automated Identification of Onset From Auroral Data

In order to address the need for a quantitative measure of auroral onset, *Murphy et al.* [2014] developed a novel technique for identifying the time interval encompassing substorm onset by maximizing a so-called "brightening factor" calculated from Fourier analysis of the product of auroral intensity and change in auroral intensity from ground-based all-sky cameras. By iteratively applying this technique to successively smaller areas of the auroral images, this technique also identifies the location of the auroral brightening. This technique thus provides an unbiased determination of the time and location of an auroral brightening but requires a predetermined list of possible onset intervals.

### 2.3. Identification of Substorm Onsets From Ground Magnetometer Data

Substorm onsets can be identified with the start of negative *H* component bays or the start of exponential growth of ULF wave power in ground-based magnetometer data collected from auroral latitudes or auroral indices. *Hsu and McPherron* [2012] visually inspected 7 h intervals of *AL* data to identify substorms. Despite a number of criteria for the identification of substorms being listed, such a visual inspection is ultimately subjective, similar to the identification of auroral onsets. *Newell and Gjerloev* [2011] used a rate of change in SML (−15 nT/min over at least 3 min) to indicate substorm onset. *Chu et al.* [2015] used intervals in which a midlatitude positive bay index peaked above 25 $nT^2$ to indicate substorm times. Both *Newell and Gjerloev* [2011] and *Chu et al.* [2015] found that their onset lists had a good agreement with auroral onsets determined from a space-based auroral imager by *Liou et al.* [2001]. *Milling et al.* [2008] and *Murphy et al.* [2009a] developed a technique for identifying substorm onset by determining when the magnetic ULF wave power increased above "the mean plus two standard deviations of the quiet time ULF wave power" for auroral zone magnetometers and determined that the onset of exponential ULF wave growth occurred several minutes before visually identified global auroral substorm onset as determined from global auroral imaging and at the start of auroral arc brightening seen in all-sky camera data [*Rae et al.*, 2009].

The above substorm onset identifications provide no further information about substorm phases other than the time of the expansion phase onset. *Juusola et al.* [2011] used the median positive and negative changes in *AL* as a basis for determining expansion and recovery phase intervals. The statistics of the substorms identified by this technique, such as the median phase lengths and total durations, were examined by *Partamies et al.* [2013]. *Forsyth et al.* [2014] applied a similar technique to the SML data. None of these studies specifically compared the calculated onset times with previously published lists of substorm onset. One notable difference between these techniques is that *Juusola et al.* [2011] only considered intervals in which the interplanetary magnetic field (IMF) had a southward component to be growth phase intervals, whereas *Forsyth et al.* [2014] took all intervals outside the expansion and recovery phases to be growth phase. This was based on the justification that solar wind coupling functions, such as the $\varepsilon$ function [*Perreault and Akasofu*, 1978], are nonzero for all but purely northward IMF; thus, energy is being added to the system at all times.

## 3. Substorm Onsets and Phases From Indices of the Electrojet (SOPHIE)

In the following, we describe a new "expert system" for identifying substorm onsets as well as the times of each substorm phase. This technique has been developed using the 1 min cadence SuperMAG AL (SML) data set, although in principle it could be applied to any auroral zone magnetic index or ground magnetometer time series.

Since there is no clear threshold value beyond which the data can be said to be indicative of a substorm (see Introduction), we identify substorms in a nonparametric manner on the basis of exceedance of a percentile in the rate of change of SML. We assume that negative changes in SML beyond a user-specified percentile level are indicative of substorm expansion phases and positive changes in SML are due to substorm recovery phases. We do not insist that a recovery phase must follow an expansion phase, since some expansion phases may lead into events such as steady magnetospheric convection (SMC) [*Sergeev et al.*, 1996; *Kissinger et al.*, 2012; *Walach and Milan*, 2015], although we modify the percentile threshold of positive changes in SML that identify the recovery phase to provide nearly equal numbers of expansion and recovery phases. We note that the average occurrence rate of SMCs is approximately 1/10 of that of substorms [*Kissinger et al.*, 2012].

*McPherron* [1970] identified the substorm growth phase as "significant deviations" away from a "quiet trace" of the *H* component from auroral zone magnetometers, although they note that the start of the growth phase is dependent on the definitions of a significant deviation and the quiet trace. Superposed epoch analysis of *AL* around substorm onset shows that, on average, *AL* shows a shallow downward trend prior to onset [e.g., *Weimer*, 1994]. However, on a case-by-case basis this signature is not always apparent, indicating that either substorms are not necessarily preceded by a growth phase or that growth phases do not have a unique signature in these data. Subsequent studies have chosen to take periods of southward IMF prior to an expansion phase onset as the growth phase [e.g., *Gjerloev et al.*, 2003; *Juusola et al.*, 2011; *Li et al.*, 2013]. However, *Petrukovich* [2000] showed that, particularly for small substorms, growth phase signatures were observed in the magnetotail even when the IMF was weakly northward. More recently, *Forsyth et al.* [2014] considered all nonexpansion and nonrecovery times to be growth phases, arguing that solar wind coupling functions such as the $\varepsilon$ function and others [see *Milan et al.*, 2012] are nonzero for all but purely northward IMF. However, solar wind energy input is not a sole indicator of a substorm growth phase, since without a measure for energy loss we cannot determine the net energy gained by the system. As such, we choose to define nonexpansion and recovery phase times as possible growth phases.

Based on the above, substorm phases are identified in a three-stage process (a flow diagram of each of these procedures is presented in the supporting information). In the first stage, substorm phases are identified by

1. low-pass filtering the data with a 30 min cutoff to remove the effect of ULF waves, commonly seen around substorm onset (see Rae and Watt, submitted, and references therein) and in the recovery phase [*Jorgensen et al.*, 1999];
2. calculating the time derivative of SML (dSML/dt) using a three-point Lagrangian interpolation;
3. calculating the percentiles of dSML/dt < 0 (expansion percentiles, EPs) and dSML/dt > 0 (recovery percentiles, RPs);
4. where dSML/dt is negative and |dSML/dt| is greater than a specified EP threshold (EPT), identifying the time as "expansion phase";
5. where dSML/dt is positive and greater than a specified RP threshold (RPT), identifying the time as "recovery phase";
6. identifying all other intervals as "possible growth phase."

As a result of using different thresholds for identifying the expansion and recovery phases, the above procedure results in short intervals of possible growth phases between expansion and recovery phases. Similarly, short intervals of possible growth phases occur between recovery and expansion phases during substorms with one or more intensifications. In the second stage, we remove these by

1. Identifying times when an expansion phase changes into a possible growth phase;
2. For each of these intervals, determine whether there is a recovery phase up to 30 min after the end of the expansion phase;
3. If a recovery phase begins within this 30 min window, find the minimum SML between the expansion and recovery phases;
4. Identify data prior to the minimum in SML as expansion phase and data following the minimum in SML as recovery phase.

A similar procedure is used to remove short possible growth phases (<30 min) between recovery and expansion phases using the local maximum in SML to separate the recovery and possible growth phases. Furthermore, we remove the following in order:

1. short (<10 min) expansion phases that occur between two possible growth phases;
2. short (<10 min) recovery phases that occur between two possible growth phases;
3. short (<30 min) possible growth phases that occur between two expansion phases;
4. short (<30 min) possible growth phases that occur between two recovery phases;
5. short (<30 min) recovery phases that occur between two possible growth phases;
6. short (<30 min) recovery phases that occur between possible growth phases and expansion phases.

Finally, since filtering the data smoothes out sharp decreases in SML, we adjust the expansion phase onset times to be at the first time at which two successive data points of the unfiltered dSML/dt are less than EPT up to 20 min after the previously determined onset time in order to account for the Gibbs phenomenon [*Gibbs*, 1898, 1899] that expands in time any sharp changes in the unfiltered data.

While EPT and RPT can be set arbitrarily and independently, we assume that we have correctly identified the expansion phases and that it logically follows that there are an equal number of expansion and recovery phases. Thus, in the third stage, we iteratively modify RPT, with EPT remaining fixed, to minimize the difference between the number of expansion and recovery phase onsets identified.

During substorms, enhancements in the eastward and westward electrojets, and their associated enhancements in the *AU* and *AL* indices, are essentially independent [*Rostoker*, 1972]. At other times, *AU* and *AL* can vary in tandem, indicating enhancements in magnetosphetic convection and thus in the global current system. We expect SML and SMU to act similarly; thus, we flag those expansion phases in which the mean or median value of |dSML|/|dSMU| over the expansion phase is less than two as potentially falsely identified substorms. We also flag the following recovery phase. For completeness, the phase identification (expansion or recovery) is retained.

The supporting information provides a list of the start times of possible growth phases (phase = 1), expansion phases (phase = 2), and recovery phases (phase = 3), along with flags indicating whether the event is an enhanced convection event (flag = 1) or not (flag = 0). Three lists for EPTs of 50%, 75%, and 90% are provided, based on SML data between 1 January 1996 to 31 December 2014.

In the following, we apply this technique to SML data from 1 January 1996 to 31 December 2014, when there were ~100 stations from which SML was derived. (Figure 1a shows the number of SML stations available over time.) We apply the technique to each year individually to minimize any solar cycle effects [*Ahn et al.*, 2000; *Tanskanen et al.*, 2002, 2011]. Figure 1 shows an example of the data processing and phase identification results from a representative day on 10 May 2005. Figure 1b shows the input data, and Figures 1c and 1d show the data processing in stage 1. Figure 1e shows the phase identifications at the end of stage 1 (possible growth phase in green, expansion in blue, and recovery in red), and Figure 1f shows the phase identifications at the end of stage 2. Figure 1 shows that Substorm Onsets and Phases from Indices of the Electrojet (SOPHIE) is able to identify the various substorm phases and that stage 2 corrects the classification of a number of data points initially labeled as possible growth phases. In the following, we test the validity of this technique through comparing the substorm onsets determined by SOPHIE with previously published lists of substorm onsets.

### 3.1. Comparison With Previous Techniques

As noted above, existing published lists provide the times of substorm expansion phase onsets only. Using SOPHIE, we identify any expansion phase onset as the time at which the phase changes into an expansion phase. In Figure 2, we compare these onsets with event lists from (a) the IMAGE spacecraft [*Frey et al.* [2004]; *Frey and Mende* [2006], black, hereafter FM06], (b) the Polar spacecraft [*Liou et al.* [2001], blue, hereafter L01], (c) the Time History of Events and Macroscale Interactions during Substorms ground-based all-sky imagers [*Nishimura et al.* [2010], green, hereafter N10], and (d) SuperMAG based on the technique described in *Newell and Gjerloev* [2011] as applied to the updated SuperMAG data covering 1996–2014 used in this study (red, hereafter NG11 for brevity). Figure 2 shows (left-hand column) the probability distribution of time differences ($\Delta t$) between onsets in these previously published lists and the closest onset from SOPHIE. Comparison between the given list and the 50% EPT onsets are shown in black, with the 75% EPT onsets shown in blue, and with the 90% onsets shown in green. Comparisons between the given lists and NG11 are shown in red. Negative $\Delta t$ indicates that onsets identified using SOPHIE occur before onsets in these lists. The right-hand column of Figure 2 shows the cumulative probability of the $|\Delta t|$. We examine the closest SOPHIE events to the events in these lists (i.e., L01, FM06, N10, and NG11 are the fiducial lists), rather than vice versa, since none of the auroral lists provide full time coverage. As such, there are events in the SOPHIE lists that cannot have a counterpart in the auroral event lists because there were no data to determine whether there was an auroral signature or not; thus, for a consistent comparison with all lists there is only a one-way correlation.

The left-hand column of Figure 2 shows that for $-30 < \Delta t < 30$ min the distribution of the time differences are sharply peaked between −2 and +3 min, with the distribution for the EPT of 75% lists giving the largest peak, and drop off rapidly away from this peak. The full width at half maximum (FWHM) of the distributions varied between 4 min for EPT of 50% up to 13 min for EPT of 90%. The lowest of these is comparable to the FWHM of the Gaussian fit to the $\Delta t$ distribution from *Chu et al.* [2015]. The SOPHIE probabilities are somewhat higher for positive $\Delta t$ for the comparisons with FM06, L01, and N10, showing that it is more likely that the onsets from SOPHIE follow auroral onsets. The $\Delta t$ distributions from the comparison with NG11 differ significantly from the other results, with a higher probability of a SOPHIE onset preceding an onset from NG11 and much higher peak probabilities of 12%, 20%, and 17% for EPT of 50%, 75%, and 90%, respectively. This results from using the same data set as NG11 but with a different approach to identifying onset. We note that the NG11 onset threshold (−15 nT/min over at least 3 min) corresponds to the 82nd percentile of the unfiltered data set. The relative excess of events between −15 min and 0 min for EPTs of 50% and 75% can be explained by considering that the dSML/dt may decrease over a few minutes prior to reaching a rate of −15 nT/min, depending on the sharpness of the change in SML. If the SML profile is such that onset is detected by both methods, the lower thresholds will be met earlier; thus, the SOPHIE onsets will precede those in NG11.

The probabilities in the left-hand columns are integrated with respect to $|\Delta t|$ to give the cumulative probabilities shown in right-hand column of Figure 2. These show the probability of a SOPHIE event being associated with an event in the comparison list (FM06, L01, N10, or NG11) within the time $|\Delta t|$ or less. This probability is higher for EPTs of 50% and 75% than EPT of 90%. Similarly, the probability of an NG11 event being associated with an event in the FM06, L01, and N10 lists within a given time frame is lower than for events from SOPHIE. We find that for EPT of 50% or 75%, half of the onsets within the tested lists are associated with a SOPHIE onset within 20 min, whereas this time > 30 min when comparing NG11 with FM06, L01, and N10. As such, using EPT of between 50% and 75%, our technique returns onset times more closely correlated to FM06, L01, and N10 than NG11. Comparing the NG11 onsets with those from SOPHIE, we find that the probability of a SOPHIE onset being associated with a NG11 onset is higher at lower $\Delta t$ than for the comparison of SOPHIE or NG11 with FM06, L01, and N10. Figure 2d shows that for EPT of 50%, 75%, and 90%, SOPHIE returns 20%, 33%, and 36% of its onsets within 1 min of onsets in the NG11 list. This is to be expected, given that the same data set and broadly similar techniques were used to determine the onset times. We are unable to directly compare FM06, L01, and N10 in a similar manner since these lists of onsets do not overlap in time.

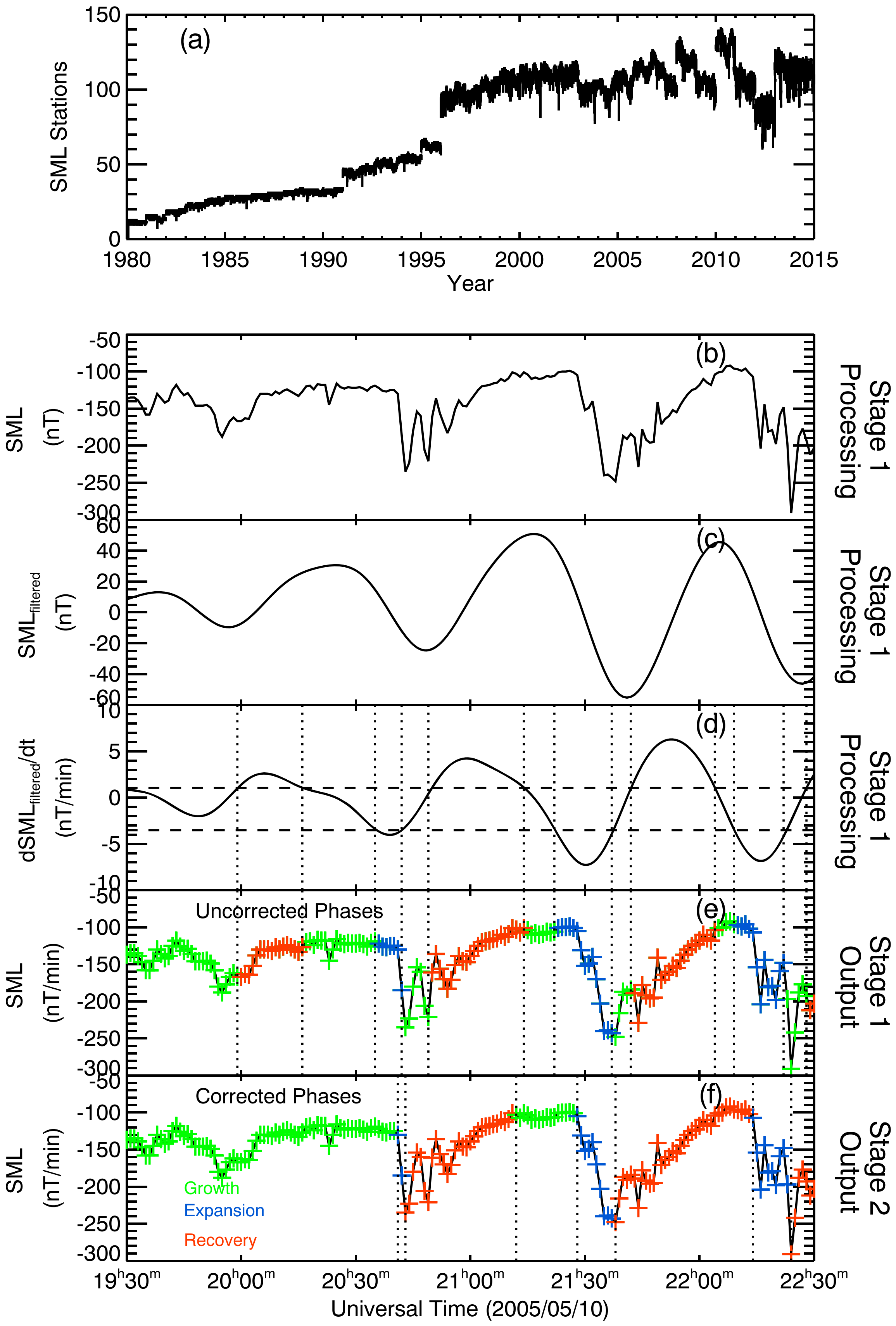

**Figure 1.** (a) The number of stations used to determine SML over time. (b–f) An example of the determination of substorm phases by SOPHIE for a 3 h period from 10 May 2005, 19:30 UT. (b) Unfiltered SML data. (c) SML data filtered using a 30 min low-pass filter. (d) The time derivative of the filtered SML data calculated using a three-point Lagrangian interpolation, with horizontal lines showing threshold dSML/dt. The vertical dotted lines show where dSML/dt crosses these thresholds. When dSML/dt is below the lower dashed line we identify the expansion phase, and when dSML/dt is above the upper dashed line we identify the recovery phase. (e) The unfiltered SML data color coded by the stage 1 phase identification (growth/energy input phase in green, expansion phase in blue, and recovery phase in red). (f) The unfiltered data color coded by the stage 2 phase identification. The vertical dotted lines indicate times when the substorm phase changes.

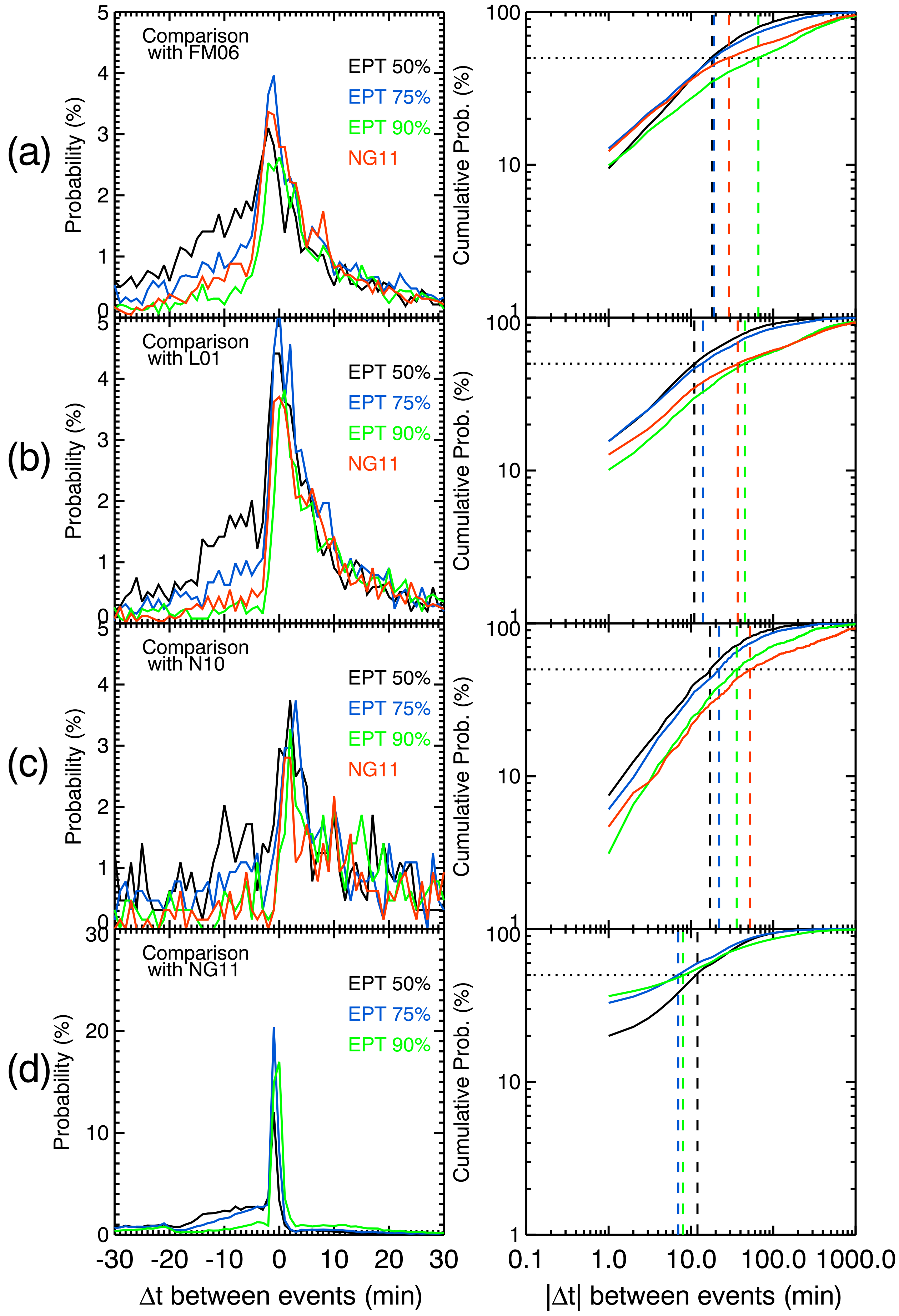

**Figure 2.** (left-hand column) Percentage probabilities and (right-hand column) cumulative probabilities of the time differences between onsets in (a) the *Frey and Mende* [2006] list, (b) the *Liou et al.* [2001] list, (c) the *Nishimura et al.* [2010] list, and (d) the *Newell and Gjerloev* [2011] list and onsets determined by SOPHIE with different EPT (black: 50%, blue: 75%, and green: 90%). A negative time difference indicates that the events from SOPHIE are listed before the corresponding event in the compared list. The horizontal line in the right-hand column indicates a cumulative probability of 50%, and the vertical color-coded lines show the time differences at which the different traces cross the 50% line.

The time for the 50% cumulative probability from a comparison of NG11 with L01 differs from that presented in NG11 because that study only compared onsets with those from L01 identified between 1997 and 1998, only considered events with a positive $\Delta t$, and discounted any events with a $\Delta t$ greater than 60 min. In our analysis, we include all events.

**Table 1.** Table Showing Some Statistics of the Output of SOPHIE for SML Data Between 1996 and 2014[a]

| EPT | | 50 | 75 | 90 |
|---|---|---|---|---|
| RPT range | | 31–41 | 59–63 | 74–82 |
| Percentage of time in substorm phases (%) | Possible growth | 36.6 | 63.8 | 84.9 |
| | Expansion | 15.5 | 9.6 | 4.1 |
| | Recovery | 26.2 | 19.3 | 9.3 |
| | Enhanced convection (expansion + recovery) | 21.7 | 7.3 | 1.7 |
| Number of phase onsets (column total corresponds number of data entries in supporting information files) | Possible growth | 31,341 | 35,772 | 24,312 |
| | Expansion | 66,554 | 45,917 | 23,846 |
| | Recovery | 67,700 | 46,593 | 24,075 |
| | Enhanced convection (expansion + recovery) | 76,590 | 25,041 | 6,208 |
| Number of transitions between substorm phases | Possible growth to expansion | 18,661 | 26,295 | 19,079 |
| | Possible growth to enhanced convection | 10,106 | 6,240 | 2,312 |
| | Recovery to expansion | 32,706 | 15,644 | 4,171 |
| | Recovery to enhanced convection | 16,164 | 4,560 | 715 |
| | Possible growth to recovery | 2,573 | 3,236 | 2,920 |
| | Expansion to possible growth | 1,427 | 2,560 | 2,691 |
| Mean phase length | Possible growth | 116.8 | 178.2 | 348.9 |
| | Expansion | 23.3 | 20.9 | 17.2 |
| | Recovery | 38.7 | 41.3 | 38.6 |
| | Enhanced convection (expansion + recovery) | 55.7 | 56.9 | 53.0 |
| Median phase length | Possible growth | 57 | 78 | 141 |
| | Expansion | 19 | 17 | 14 |
| | Recovery | 30 | 34 | 33 |
| | Enhanced convection (expansion + recovery) | 48 | 49 | 47 |
| Waiting times between isolated expansion phase onsets with less than 10 h separation (hours) | Median | 3.92 | 3.70 | 3.72 |
| | Mean | 4.33 | 4.13 | 4.18 |
| | Standard deviation | 2.50 | 2.35 | 2.35 |
| Mean onsets (possible growth to expansion) per day | | 4.18 | 4.78 | 3.12 |
| Mean intensifications per day | | 11.03 | 3.73 | 0.82 |

[a]Shown are the proportions of the data set identified as growth, expansion, and recovery phases; the number of growth, expansion, and recovery phase onsets; the number of transitions between the various phases; and measures of the distribution of waiting times between expansion phase onsets preceded by a possible growth phase. The mean number of expansion phase onsets preceded by a growth phase or preceded by a recovery phase (intensifications) are shown in the bottom rows. The first row shows the EPT value tested, and the second row shows the range of RPT values obtained.

Table 1 shows the proportion of time taken up by each phase, the number of phase onsets, and the number of transitions between different phases. Those expansion phases in which the difference in the variation of SML and SMU is low, along with any subsequent recovery phases, are listed as "enhanced convection." Also shown are the means and medians of the phase lengths and the waiting times between expansion phase onsets directly preceded by a possible growth phase.

For higher values of EPT (corresponding to faster negative changes in SML), the proportion of the data marked as growth phase increases and the proportion marked as expansion and recovery correspondingly decreases. The percentage of time in the recovery phase for list from EPT of 50%, 75%, and 90% is almost double that in the expansion phase. The mean and median phase lengths of the recovery phases are approximately constant with changing EPT, but the expansion phase durations shrink with increasing EPT. Our average recovery phase lengths are consistent with the values from *Partamies et al.* [2013], although our mean expansion phase lengths are slightly shorter and our median expansion phase lengths are slightly longer. The mean and median lengths of the possible growth phase intervals from SOPHIE are greater than the growth phase lengths from *Partamies et al.* [2013] and *McPherron* [1994]; however, this is to be expected as they used a more restrictive criterion for identifying the growth phase.

For higher EPT values, there are fewer substorm intensifications (expansion phases following a recovery phase interval) than expansion phases following a possible growth phase, suggesting that intensifications are relatively small. One of the benefits of SOPHIE over the existing event lists examined here is that our technique enables the explicit determination of whether an expansion phase onset is a substorm intensification.

For an EPT of 75, SOPHIE identifies 26,295 substorms following a possible growth phase, within 13% of the 30,484 substorms identified by NG11 in the same period, although the total number of expansion phases is higher (45,917, 50% more than NG11). At lower EPT SOPHIE identifies more expansion phases than NG11, and at EPT of 90 SOPHIE identifies fewer expansion phases than NG11. A comparison with the number of onsets in FM06, N10, and L01 is complicated due to the incomplete time coverage of the instruments used in these studies.

Figure 3 shows the waiting time distributions between (a) substorms identified in the FM06, L01, N10, and NG11 lists; (b) substorms from SOPHIE preceded by a possible growth phase for EPT values of 50%, 75%, and 90%; and (c) all the expansion phase onsets from SOPHIE (including substorm intensifications) for the given EPT values. The waiting time distributions from each of the previously published lists are markedly different, despite all apparently determining substorm onsets. While this may be a result of the times covered by these lists and the variations in substorm activity with solar cycle, it is more likely that these differences result from the different techniques used in creating these lists. NG11, L01, and N10 all show peaks in the distributions at ~30 min, whereas FM06 peaks at ~60 min. The waiting time distributions from onsets identified by SOPHIE preceded by a possible growth phase (Figure 3b) are similar to that from FM06 (Figure 3a), whereas the distributions of all expansion phase onsets from SOPHIE (Figure 3c) are more similar to NG11, L01, and N10 (Figure 3a). This suggests that a large number of events in the NG11, L01, and N10 lists are in fact intensifications.

While SOPHIE gives users the ability to determine onset times using any EPT, based on our analysis above, we recommend that an EPT of 75% to give a list of onsets similar to previous studies.

### 3.2. Superposed Epoch Analysis of Substorm SML

In order to examine whether or not the general profile of the substorms identified by SOPHIE is consistent with those from previously published onset list, we compare the results of a superposed epoch analysis of SML. Figure 4 shows the results of this analysis with the zero epoch defined as the expansion phase onset times from the auroral and magnetic lists and from SOPHIE, using an EPT of 75%. For each trace, we set the median SML value at zero epoch to 0 nT. Data from FM06 onsets are shown in black, L01 are shown in blue, N10 are shown in yellow, and NG11 in red. Data from SOPHIE (EPT of 75%) including all expansion phase onsets are shown as a dotted black line, and results only onsets following a possible growth phase are shown as the dashed black line.

The superposed epoch study results based on the auroral lists (blue, green, and solid black lines) generally show similar trends: SML is approximately constant prior to onset, then SML rapidly decreases over ~25 min, before recovering to its preonset level over the next 150 min. Results from the N10 list are different, however, and give a minimum in SML at 57 min after onset. Figure 3 demonstrates that the tested substorm onset lists generally show the same large-scale characteristics apart from N10. While the N10 events show a small geomagnetic bay following their onsets, it is much weaker and shows no sharp decrease in SML at the onset time. This suggests that the auroral intensifications identified by N10 are most likely intensifications of a preexisting current system rather than a substorm onset, as first pointed out by *Frey* [2010].

The superposed epoch results from the events from the magnetically derived SOPHIE and NG11 lists show similar trends, although the magnitude of the events in NG11 is approximately double that of the events from SOPHIE. Unlike the auroral lists, the traces from SOPHIE including all onsets (dotted line) and from NG11 show a distinct minimum ~10 min prior to onset. This minimum is removed if only events with onsets following a growth phase are used (dashed line). This shows that the local minimum in the superposed epoch analysis of SML prior to onset is due to the identification of substorm intensifications as onsets and strongly suggests that some of the onsets in NG11 are in fact substorm intensifications.

In summary, onsets from SOPHIE are generally identified prior to auroral onsets with more than 50% of auroral onsets occurring within 20 min of an onset identified by SOPHIE. The distributions of the time differences between onsets from SOPHIE are similar in shape to those found by comparing NG11 to auroral onset lists. The average length of the expansion and recovery phases determined by SOPHIE is comparable with previous studies, but the possible growth phase is much longer due to a much less restrictive definition of this phase. Comparing the superposed epoch analysis of SML determined using each list shows similar trends.

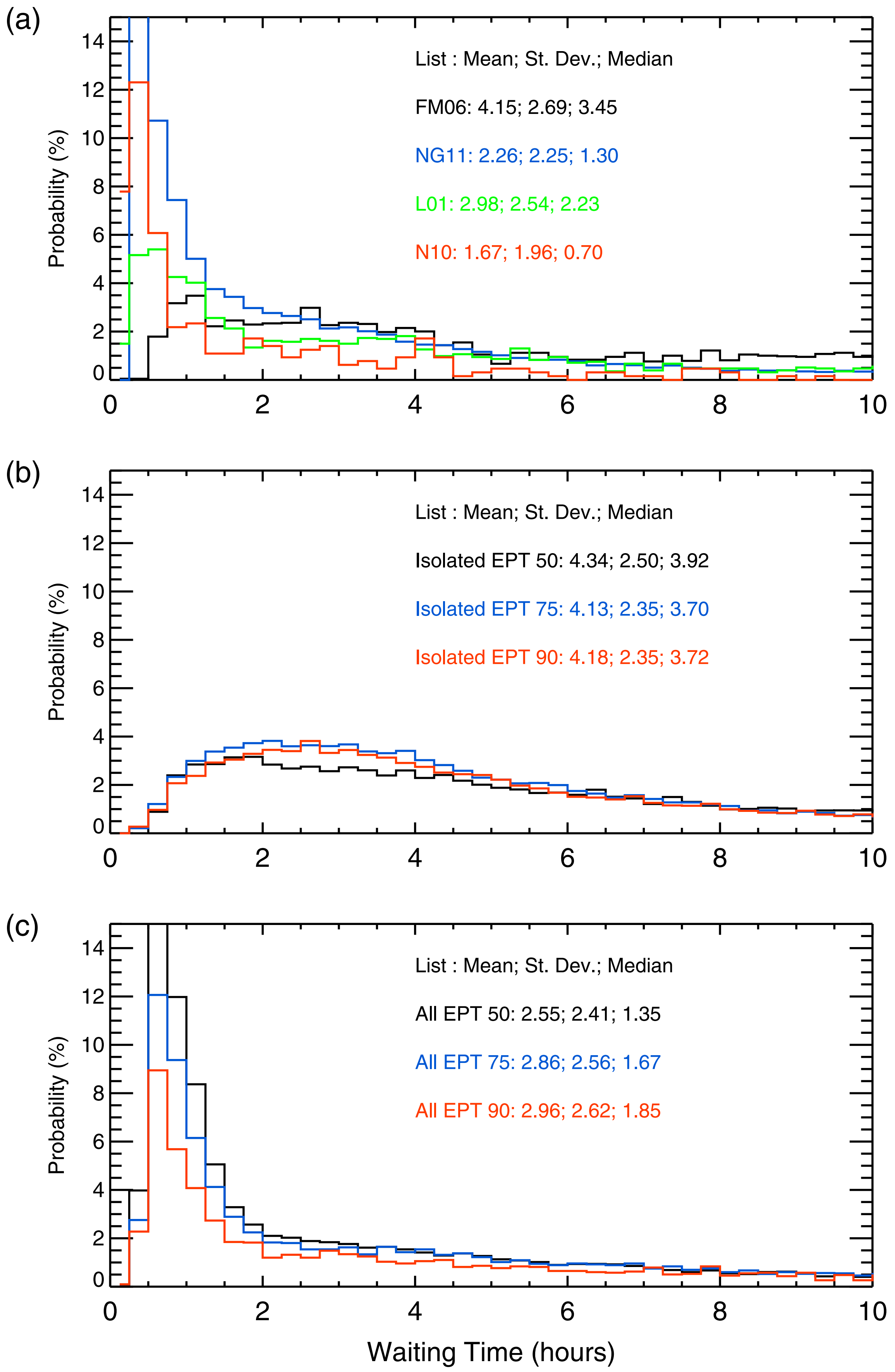

**Figure 3.** The distribution of waiting times between substorm expansion phases onsets for (a) the (black) *Frey and Mende* [2006], (blue) *Newell and Gjerloev* [2011], (green) *Liou et al.* [2001], and (red) *Nishimura et al.* [2010] lists; (b) isolated (i.e., preceded by a possible growth phase) substorms from SOPHIE; and (c) all substorm onsets from SOPHIE. In Figures 3b and 3c, black shows onsets determined with an EPT of 50%, blue shows onsets for an EPT of 75%, and red shows onsets for an EPT of 90%. The mean, median, and standard deviations of the displayed distributions are shown next to the list identifiers.

## 4. Discussion

We have developed a novel technique, called SOPHIE, for identifying substorms from auroral electrojet indices. This new technique enables us to identify the start and end of each phase, including substorm intensifications, thus providing a complete picture of substorm activity, as opposed to solely identifying the expansion phase onset. Furthermore, by using magnetometer-based auroral indices, SOPHIE provides a near-continuous, long-term identification of substorm phases. We have shown that the expansion phase onset times from SOPHIE are similar to, although often earlier than, those from existing automated and visually determined lists.

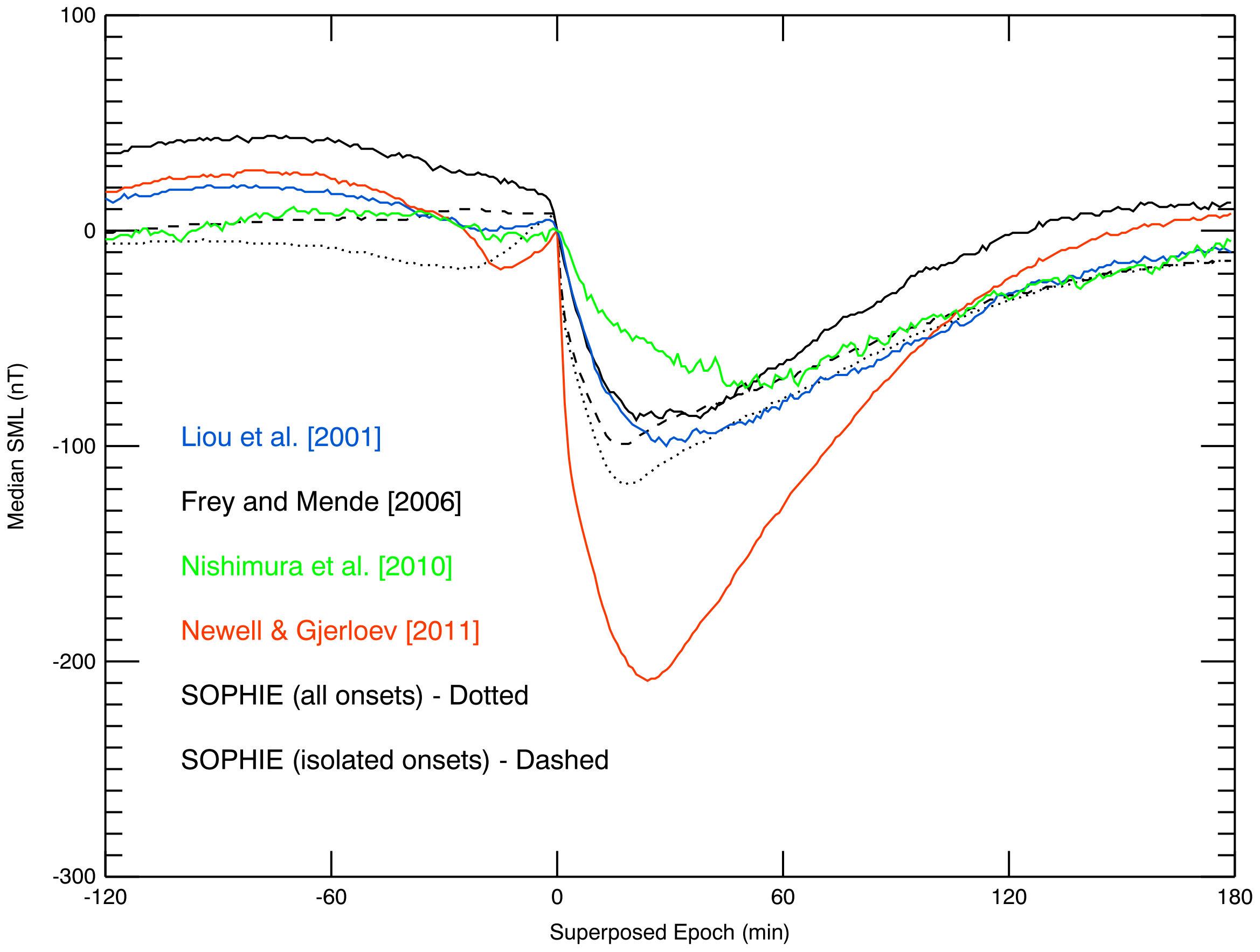

**Figure 4.** Superposed epoch analysis of the SuperMAG SML index using onsets from four established substorm onset lists and from SOPHIE. Zero epoch is set at the expansion phase onset time. The median SML value at zero epoch has been removed from each trace. Superposed epoch analyses of SML using onsets from *Liou et al.* [2001] (blue), *Frey et al.* [2004] (black), *Nishimura et al.* [2010] (green), and *Newell and Gjerloev* [2011] (red) are shown along with the analysis of all expansion phase onsets from SOPHIE (dotted line) and isolated expansion phase onsets from SOPHIE (dashed line). The trends in the superposed data from each list are similar, although the magnitudes of the variations differ. We note that a slight upturn in SML prior to onset is seen from the *Newell and Gjerloev* [2011] list and from the SOPHIE onsets including compound substorms but is not seen in the data from the auroral lists or isolated onsets from SOPHIE.

SOPHIE defines the substorm expansion or recovery phase based on exceedance of some threshold in the rate of change of SML. Given that the distribution of dSML/dt appears continuous, there is no obvious threshold value to choose. Consequently, SOPHIE uses a nonparametric approach (i.e., not based on physical units), defining substorm times based on a given percentile of the rate of change of SML. In this way, the method is insensitive to changes in SML, such as seasonal variations [*Singh et al.*, 2013], solar cycle variations [*Ahn et al.*, 2000; *Tanskanen et al.*, 2002, 2011] and changes in the number of stations in the magnetometer network, in the sense that the proportion of expansion and recovery phases identified is essentially preserved for given EPT and RPT values. More precisely, an EPT of 90% will pick out the largest 10% of negative changes in SML as expansion phase events. Smaller EPT and RPT limits will enable smaller changes in SML to be identified as substorms and thus will pick out smaller events. It should be noted that this would also move the expansion phase onsets of large events to earlier times. Also, it could be argued that the optimal EPT and RPT values are those that yield equal number of expansion and recovery phase intervals. Our method iteratively calculates the RPT value (to within 1%) that minimizes the difference between the number of expansion and recovery phases.

With this technique, we present a means to provide an accurate and continuous and, most importantly, user-independent determination of substorm onset, as well as substorm expansion, recovery, and potential growth phase durations. As with all automated procedures, it is necessary to calibrate against existing results. We have calibrated the output of SOPHIE against the substorm onset lists of *Liou et al.* [2001], *Frey and Mende* [2006], *Nishimura et al.* [2010], and *Newell and Gjerloev* [2011] as well as with the reported results of other studies. We provide the results of three different EPT values and find that an EPT of 75% renders expansion phase onsets within 20 min of those in other lists and that the superposed epoch analyses of SML using the 75% EPT were similar to superposed epoch analysis results of the above auroral onset lists. As such, we recommend that an EPT of 75% be used to identify substorms similar to those within the literature, although we also include both the 50% and 90% EPT values to enable researchers to study small, medium, and large substorm events.

We note that SOPHIE tends to identify expansion phase onset times after onsets determined from auroral observations. A similar trend was seen by *Chu et al.* [2015], who identified substorm onsets from midlatitude magnetometer data. This is most likely due to requiring a sufficiently high threshold to exclude small-scale variations that are not substorms. We note that recent studies have shown that ULF wave activity begins prior to auroral brightenings observed by global auroral imagers [*Murphy et al.*, 2009a] but contemporaneous with ground-based all-sky imager observations of the brightening of the substorm arc [*Rae et al.*, 2009]. The formation of the geomagnetic bay may occur prior to the time at which the aurorae are sufficiently bright over a large enough area for a global auroral imager to measure a discernable change in auroral brightness. We thus conclude that our onset times are as valid as other substorm onset times, although while the mechanisms behind onset remain unclear, determining true onset times is impossible.

Unlike the most commonly used onset lists, SOPHIE is not limited to identifying expansion phase onset; we also identify the start of the possible growth phases and recovery phases and consequently the durations of all substorm phases. A comparison between average phase lengths and substorm waiting times shows that SOPHIE gives comparable results to earlier studies [*Borovsky et al.*, 1993; *Hsu and McPherron*, 2012; *Partamies et al.*, 2013], with the notable exception that our possible growth phases are somewhat longer due to our less restrictive criteria in identifying these phases.

The superposed epoch analysis (Figure 4) shows that while the overall trends in SML for the events identified in each of these lists are similar, there are some interesting differences: notably, the magnitude of the bays from *Newell and Gjerloev* [2011] were approximately twice that from the auroral lists, the minimum in SML from the *Nishimura et al.* [2010] list was 25 min later than the other lists, and the SML traces from all the events identified by SOPHIE and from *Newell and Gjerloev* [2011] showed an increase in SML just prior to the expansion phase onset. The onsets from *Liou et al.* [2001] and *Nishimura et al.* [2010] were identified close to or during solar minima, whereas events from *Frey and Mende* [2006] and *Nishimura et al.* [2010] were identified close to solar maximum. Larger substorms are typically seen following solar maximum [*Tanskanen et al.*, 2011]; thus, this may account for the difference in the median SML for the events from these auroral lists. The larger magnitude of the substorm bays from the *Newell and Gjerloev* [2011] list is likely due to a threshold which picks out changes in SML above the 82nd percentile. Using a lower percentile will still identify these large events but also identify smaller events, as the auroral lists do. The upturn in SML just prior to onset seen in the superposed epoch analysis of all the SOPHIE events and the *Newell and Gjerloev* [2011] events is due to the identification of substorm intensifications (expansion phases immediately following a recovery phase). The studied auroral lists all discount further brightening of the aurora within a specified time from an identified onset; thus, they tend not to identify subsequent substorm intensifications. This is demonstrated by isolating those events found by SOPHIE that are preceded by a possible growth phase interval (dotted line in Figure 4), which shows only a slight upturn in SML prior to onset. Such a comparison has not previously been possible, as the existing lists do not identify each substorm phase.

Like previous techniques, SOPHIE is limited in that it distils substorm processes into a set of signatures in a single data set. Previous studies have shown that combining substorm signatures from the aurora with the occurrence of particle injections at geosynchronous orbit introduces uncertainty into the identification of substorms [*Boakes et al.*, 2011]. From the waiting time distributions presented here and in previous studies [*Borovsky et al.*, 1993; *McPherron*, 1994; *Hsu and McPherron*, 2012], we can see that different techniques designed to identify the same global phenomenon determine events with very different temporal distributions as a result of using different techniques and data sets. In practice, SOPHIE identifies times of increasing or decreasing deflection of the ground magnetic field at auroral latitudes, which can be linked to increase or decrease in the strength of the westward electrojet. This is commonly reported as a signature of expansion and recovery phase activity as a so-called "negative bay" [*Davis and Sugiura*, 1966]. However, there are a zoo of events that are not substorms but that may exhibit some substorm-like features: pseudobreakups and convection bays may be identified as expansion and recovery phase times and steady magnetospheric convection events may be identified as possible growth phases. Users of this technique should be aware of these limitations and that using SOPHIE in conjunction with observations of auroral activity, wave activity, and particle injections will give a more comprehensive list of substorm events. However, by using SOPHIE we can improve our knowledge of the processes within the substorm cycle by comparing data not only at times before or after expansion phase onset but at specific intervals within the possible growth, expansion, and recovery phases.

Existing lists of substorms concentrate solely on substorm onset, either by identifying auroral brightenings or sharp decreases in auroral indices. These are often used as the basis of statistical superposed epoch analysis of a variety of different data sets. However, given that substorm phases can vary in length, the validity of comparing data at times increasingly far from the defined onset reduces as the analysis mixes growth, expansion, and recovery phase data. *Yokoyama and Kamide* [1997] and *Hutchinson et al.* [2011] discussed this shortcoming of superposed epoch analysis in terms of geomagnetic storms. Using SOPHIE, we identify the start and duration of each substorm phase, therefore enabling future studies to perform superposed epoch analyses of individual phases or comparing data points within given fractions of the total substorm or substorm phases.

While we have applied this technique to the SML data set, in principle it could be applied to any auroral zone magnetometer or magnetometer chain. Given that the thresholds for identifying each phase are based on the variations in the data set, rather than absolute values, this may enable substorm phases to be more readily identified in stations that are typically further from the auroral electrojet. However, study of this application is beyond the scope of this technique paper.

## 5. Summary

We have developed a new technique, called SOPHIE, for identifying all substorm growth/energy input, expansion and recovery phase onsets, and durations from auroral electrojet indices and ground magnetometer data. In order to test this technique, we applied it to the SuperMAG AL (SML) index. In summary, the technique works as follows:

1. Intervals during which |dSML/dt| is greater than a specified percentile (EPT) of the dSML/dt $< 0$ data set are marked as expansion phase;
2. Intervals during which dSML/dt is greater than a specified percentile (RPT) of the dSML/dt $> 0$ data set are marked as recovery phase;
3. All other data are marked as possible growth phase.

The data are then processed to remove specifically identified artifacts. The RPT value is adjusted iteratively to minimize the difference between the number of expansion phases identified and the number of recovery phases identified.

Comparing the expansion phase onsets determined using this new technique with previously published lists, we find that at least 50% of the time there is always a SOPHIE event within 20 min of events in the comparison list, although our method tends to identify the substorm onset time after those in auroral lists. We determined that an EPT of 75% gives a comparable distribution of substorm onsets to previous lists. Through a comparison with existing lists and recent results examining the timing of various onset signatures, we conclude that our identified onsets are at least as valid as existing lists of substorm onsets. Unlike many other techniques, SOPHIE provides the opportunity for far more detailed study of individual substorm phases during substorms as a whole and not just around the expansion phase onset by identifying the time and duration of each individual substorm phase. Thus, SOPHIE is an improvement over existing lists of substorm onsets.

**Acknowledgments**
For the SuperMAG indices and ground magnetometer data, we gratefully acknowledge Intermagnet; USGS, Jeffrey J. Love; CARISMA, PI Ian Mann; CANMOS; The S-RAMP Database, PI K. Yumoto and K. Shiokawa; The SPIDR database; AARI, PI Oleg Troshichev; The MACCS program, PI M. Engebretson, Geomagnetism Unit of the Geological Survey of Canada; GIMA; MEASURE, UCLA IGPP and Florida Institute of Technology; SAMBA, PI Eftyhia Zesta; 210 Chain, PI K. Yumoto; SAMNET, PI Farideh Honary; the institutes who maintain the IMAGE magnetometer array, PI Eija Tanskanen; AUTUMN, PI Martin Connors; DTU Space, PI Jürgen Matzka; South Pole and McMurdo Magnetometer, PI's Louis J. Lanzarotti and Alan T. Weatherwax; ICESTAR; RAPIDMAG; PENGUIn; British Antarctic Survey; McMac, PI Peter Chi; BGS, PI Susan Macmillan; Pushkov Institute of Terrestrial Magnetism, Ionosphere and Radio Wave Propagation (IZMIRAN); GFZ, PI Jürgen Matzka; MFGI, PI B. Heilig; IGFPAS, PI J. Reda; University of L'Aquila, PI M. Vellante; and SuperMAG, PI Jesper W. Gjerloev. The SuperMAG SML indices and substorm onsets can be downloaded from http://supermag.jhuapl.edu. C.F. and I.J.R. are funded in part by Natural Environment Research Council (NERC) grants NE/L007495/1 and NE/M00886X/1 and IJR by Science and Technology Facilities Council (STFC) grant ST/L000563/1. C.M.J. and J.C.C. are funded by NERC grant NE/L007177/1. C.M.J. is also supported by an STFC Ernest Rutherford Fellowship. M.P.F.'s contribution is funded by NERC grant NE/L006456/1. A.N.F. is funded by STFC grants ST/K000977/1 and ST/L005638/1.

## References

Ahn, B. H., H. W. Kroehl, Y. Kamide, and E. A. Kihn (2000), Seasonal and solar cycle variations of the auroral electrojet indices, *J. Atmos. Sol. Terr. Phys.*, *62*(14), 1301–1310.
Akasofu, S. I. (1964), The development of the auroral substorm, *Planet. Space Sci.*, *12*(4), 273–282.
Akasofu, S. I., and S. Chapman (1961), Ring current, geomagnetic disturbance, and Van Allen radiation belts, *J. Geophys. Res.*, *66*(5), 1321–1350, doi:10.1029/JZ066i005p01321.
Axford, W. I. (1969), Magnetospheric convection, *Rev. Geophys.*, *7*(1–2), 421–459.
Boakes, P. D., S. E. Milan, G. A. Abel, M. P. Freeman, G. Chisham, and B. Hubert (2009), A statistical study of the open magnetic flux content of the magnetosphere at the time of substorm onset, *Geophys. Res. Lett.*, *36*, L04105, doi:10.1029/2008GL037059.

Boakes, P. D., S. E. Milan, G. A. Abel, M. P. Freeman, G. Chisham, and B. Hubert (2011), A superposed epoch investigation of the relation between magnetospheric solar wind driving and substorm dynamics with geosynchronous particle injection signatures, *J. Geophys. Res.*, *116*, A01214, doi:10.1029/2010JA016007.

Borovsky, J. E., R. J. Nemzek, and R. D. Belian (1993), The occurrence rate of magnetospheric-substorm onsets: Random and periodic substorms, *J. Geophys. Res.*, *98*(A3), 3807–3813, doi:10.1029/92JA02556.

Chu, X., R. L. McPherron, T.-S. Hsu, and V. Angelopoulos (2015), Solar cycle dependence of substorm occurrence and duration: Implications for onset, *J. Geophys. Res. Space Physics*, *120*, 2808–2818, doi:10.1002/2015JA021104.

Consolini, G., and P. De Michelis (1998), Non-Gaussian distribution function of *AE*-index fluctuations: Evidence for time intermittency, *Geophys. Res. Lett.*, *25*(21), 4087–4090, doi:10.1029/1998GL900073.

Coumans, V., C. Blockx, J. C. Gerard, B. Hubert, and M. Connors (2007), Global morphology of substorm growth phases observed by the IMAGE-SI12 imager, *J. Geophys. Res.*, *112*, A11211, doi:10.1029/2007JA012329.

Davis, T. N., and M. Sugiura (1966), Auroral electrojet activity index *AE* and its universal time variations, *J. Geophys. Res.*, *71*(*3*), 785−801, doi:10.1029/JZ071i003p00785.

Forsyth, C., et al. (2014), Increases in plasma sheet temperature with solar wind driving during substorm growth phases, *Geophys. Res. Lett.*, *41*, 8713–8721, doi:10.1002/2014GL062400.

Frey, H. U. (2010), Comment on “Substorm triggering by new plasma intrusion: THEMIS all-sky imager observations” by Y. Nishimura et al., *J. Geophys. Res.*, *115*, A12232, doi:10.1029/2010JA016113.

Frey, H. U., and S. B. Mende (2006), Substorm onsets as observed by IMAGE-FUV, paper presented at Eighth International Conference on Substorms (ICS-8), Univ. of Calgary, Alberta, Canada, 2007.

Frey, H. U., S. B. Mende, V. Angelopoulos, and E. F. Donovan (2004), Substorm onset observations by IMAGE-FUV, *J. Geophys. Res.*, *109*, A10304, doi:10.1029/2004JA010607.

Gibbs, J. W. (1898), Fourier’s series, *Nature*, *59*, 200.

Gibbs, J. W. (1899), Fourier’s series, *Nature*, *59*, 606.

Gjerloev, J. W. (2012), The SuperMAG data processing technique, *J. Geophys. Res.*, *117*, A09213, doi:10.1029/2012JA017683.

Gjerloev, J. W., R. A. Hoffman, E. Tanskanen, and M. Friel (2003), Auroral electrojet configuration during substorm growth phase, *Geophys. Res. Lett.*, *30*(18), 1927, doi:10.1029/2003GL017851.

Gjerloev, J. W., R. A. Hoffman, J. B. Sigwarth, and L. A. Frank (2007), Statistical description of the bulge-type auroral substorm in the far ultraviolet, *J. Geophys. Res.*, *112*, A07213, doi:10.1029/2006JA012189.

Heppner, J. P. (1954), Time sequences and spatial relations in auroral activity during Magnetic Bays at College, Alaska, *J. Geophys. Res.*, *59*(3), 329–338, doi:10.1029/JZ059i003p00329.

Hsu, T. S., and R. L. McPherron (2012), A statistical analysis of substorm associated tail activity, *Adv. Space Res.*, *50*(10), 1317–1343.

Hutchinson, J. A., D. M. Wright, and S. E. Milan (2011), Geomagnetic storms over the last solar cycle: A superposed epoch analysis, *J. Geophys. Res.*, *116*, A09211, doi:10.1029/2011JA016463.

Jorgensen, A. M., H. E. Spence, T. J. Hughes, and D. McDiarmid (1999), A study of omega bands and Ps6 pulsations on the ground, at low altitude and at geostationary orbit, *J. Geophys. Res.*, *104*(A7), 14,705–14,715, doi:10.1029/1998JA900100.

Juusola, L., N. Ostgaard, E. Tanskanen, N. Partamies, and K. Snekvik (2011), Earthward plasma sheet flows during substorm phases, *J. Geophys. Res.*, *116*, A10228, doi:10.1029/2011JA016852.

Kissinger, J., R. L. McPherron, T. S. Hsu, V. Angelopoulos, and X. Chu (2012), Necessity of substorm expansions in the initiation of steady magnetospheric convection, *Geophys. Res. Lett.*, *39*, L15105, doi:10.1029/2012GL052599.

Kistler, L. M., et al. (2006), Ion composition and pressure changes in storm time and nonstorm substorms in the vicinity of the near-Earth neutral line, *J. Geophys. Res.*, *111*, A11222, doi:10.1029/2006JA011939.

Li, H., C. Wang, and Z. Peng (2013), Solar wind impacts on growth phase duration and substorm intensity: A statistical approach, *J. Geophys. Res. Space Physics*, *118*, 4270–4278, doi:10.1002/jgra.50399.

Liou, K., P. T. Newell, D. G. Sibeck, C. I. Meng, M. Brittnacher, and G. Parks (2001), Observation of IMF and seasonal effects in the location of auroral substorm onset, *J. Geophys. Res.*, *106*(A4), 5799–5810, doi:10.1029/2000JA003001.

McPherron, R. L. (1970), Growth phase of magnetospheric substorms, *J. Geophys. Res.*, *75*(28), 5592–5599, doi:10.1029/JA075i028p05592.

McPherron, R. L. (1994), The growth phase of magnetospheric substorms, paper presented at Second International Conference on Substorms, Geophysical Institute, Univ. of Alaska, Fairbanks, Fairbanks, Alaska.

McPherron, R. L., C. T. Russell, and M. P. Aubry (1973), Satellite studies of magnetospheric substorms on August 15, 1968: 9. Phenomenological model for substorms, *J. Geophys. Res.*, *78*(16), 3131–3149, doi:10.1029/JA078i016p03131.

Milan, S. E., A. Grocott, C. Forsyth, S. M. Imber, P. D. Boakes, and B. Hubert (2009), A superposed epoch analysis of auroral evolution during substorm growth, onset and recovery: Open magnetic flux control of substorm intensity, *Ann. Geophys.-Germany*, *27*(2), 659–668.

Milan, S. E., J. S. Gosling, and B. Hubert (2012), Relationship between interplanetary parameters and the magnetopause reconnection rate quantified from observations of the expanding polar cap, *J. Geophys. Res.*, *117*, A03226, doi:10.1029/2011JA017082.

Milling, D. K., I. J. Rae, I. R. Mann, K. R. Murphy, A. Kale, C. T. Russell, V. Angelopoulos, and S. Mende (2008), Ionospheric localisation and expansion of long-period Pi1 pulsations at substorm onset, *Geophys. Res. Lett.*, *35*, L17S20, doi:10.1029/2008GL033672.

Murphy, K. R., I. J. Rae, I. R. Mann, D. K. Milling, C. E. J. Watt, L. Ozeke, H. U. Frey, V. Angelopoulos, and C. T. Russell (2009a), Wavelet-based ULF wave diagnosis of substorm expansion phase onset, *J. Geophys. Res.*, *114*, A00C16, doi:10.1029/2008JA013548.

Murphy, K. R., I. J. Rae, I. R. Mann, A. P. Walsh, D. K. Milling, C. E. J. Watt, L. Ozeke, H. U. Frey, V. Angelopoulos, and C. T. Russell (2009b), Reply to comment by K. Liou and Y.-L. Zhang on “Wavelet-based ULF wave diagnosis of substorm expansion phase onset”, *J. Geophys. Res.*, *114*, A10207, doi:10.1029/2009JA014351.

Murphy, K. R., D. M. Miles, C. E. J. Watt, I. J. Rae, I. R. Mann, and H. U. Frey (2014), Automated determination of auroral breakup during the substorm expansion phase using all-sky imager data, *J. Geophys. Res. Space Physics*, *119*, 1414–1427, doi:10.1002/2013JA018773.

Newell, P. T., and J. W. Gjerloev (2011), Evaluation of SuperMAG auroral electrojet indices as indicators of substorms and auroral power, *J. Geophys. Res.*, *116*, A12211, doi:10.1029/2011JA016779.

Nishimura, Y., L. Lyons, S. Zou, V. Angelopoulos, and S. Mende (2010), Substorm triggering by new plasma intrusion: THEMIS all-sky imager observations, *J. Geophys. Res.*, *115*, A07222, doi:10.1029/2009JA015166.

Opgenoorth, H. J., M. A. L. Persson, T. I. Pulkkinen, and R. J. Pellinen (1994), Recovery phase of magnetospheric substorms and its association with morning-sector aurora, *J. Geophys. Res.*, *99*(A3), 4115–4129, doi:10.1029/93JA01502.

Partamies, N., L. Juusola, E. Tanskanen, and K. Kauristie (2013), Statistical properties of substorms during different storm and solar cycle phases, *Ann. Geophys. Germany*, *31*(2), 349–358.

Perreault, P., and S. I. Akasofu (1978), Study of geomagnetic storms, *Geophys. J. R. Astron. Soc.*, *54*(3), 547–573.

Petrukovich, A. A. (2000), The growth phase: Comparison of small and large substorms, in *Proceedings of the Fifth International Conference on Substorms*, *ESA SP-443*, pp. 9–14, ESA Publications Division Noordwijk, Netherlands.

Pytte, T., R. L. Mcpherron, and S. Kokubun (1976), Ground signatures of expansion phase during multiple onset substorms, *Planet. Space Sci.*, *24*(12), 1115–1132.

Rae, I. J., et al. (2009), Timing and localization of ionospheric signatures associated with substorm expansion phase onset, *J. Geophys. Res.*, *114*, A00C09, doi:10.1029/2008JA013559.

Rae, I. J., C. E. J. Watt, K. R. Murphy, H. U. Frey, L. G. Ozeke, D. K. Milling, and I. R. Mann (2012), The correlation of ULF waves and auroral intensity before, during and after substorm expansion phase onset, *J. Geophys. Res.*, *117*, A08213, doi:10.1029/2012JA017534.

Rostoker, G. (1972), Geomagnetic indexes, *Rev. Geophys. Space. Phys.*, *10*(4), 935–950.

Sergeev, V. A., R. J. Pellinen, and T. I. Pulkkinen (1996), Steady magnetospheric convection: A review of recent results, *Space Sci. Rev.*, *75*(3–4), 551–604.

Singh, A. K., R. Rawat, and B. M. Pathan (2013), On the UT and seasonal variations of the standard and SuperMAG auroral electrojet indices, *J. Geophys. Res. Space Physics*, *118*, 5059–5067, doi:10.1002/jgra.50488.

Tanskanen, E. I., T. I. Pulkkinen, H. E. J. Koskinen, and J. A. Slavin (2002), Substorm energy budget during low and high solar activity: 1997 and 1999 compared, *J. Geophys. Res.*, *107*(A6), 1086, doi:10.1029/2001JA900153.

Tanskanen, E. I., T. I. Pulkkinen, A. Viljanen, K. Mursula, N. Partamies, and J. A. Slavin (2011), From space weather toward space climate time scales: Substorm analysis from 1993 to 2008, *J. Geophys. Res.*, *116*, A00I34, doi:10.1029/2010JA015788.

Vassiliadis, D., A. J. Klimas, D. N. Baker, and D. A. Roberts (1996), The nonlinearity of models of the vB(south)-AL coupling, *J. Geophys. Res.*, *101*(A9), 19,779–19,787, doi:10.1029/96JA01408.

Voronkov, I. O., E. F. Donovan, and J. C. Samson (2003), Observations of the phases of the substorm, *J. Geophys. Res.*, *108*(A2), 1073, doi:10.1029/2002JA009314.

Walach, M. T., and S. E. Milan (2015), Are steady magnetospheric convection events prolonged substorms?, *J. Geophys. Res. Space Physics*, *120*, 1751–1758, doi:10.1002/2014JA020631.

Weimer, D. R. (1994), Substorm time constants, *J. Geophys. Res.*, *99*(A6), 11,005–11,015, doi:10.1029/93JA02721.

Yokoyama, N., and Y. Kamide (1997), Statistical nature of geomagnetic storms, *J. Geophys. Res.*, *102*(A7), 14,215–14,222, doi:10.1029/97JA00903.